# PHYSICS-INTEGRATED MACHINE LEARNING: EMBEDDING A NEURAL NETWORK IN THE NAVIER-STOKES EQUATIONS. PART II

Arsen S. Iskhakov*, Nam T. Dinh**

*North Carolina State University, Department of Nuclear Engineering, Campus Box 7909, Raleigh, NC 27695 USA
aiskhak@ncsu.edu

**North Carolina State University, Department of Nuclear Engineering

## ABSTRACT

The work is a continuation of a paper by Iskhakov A.S. and Dinh N.T. "Physics-integrated machine learning: embedding a neural network in the Navier-Stokes equations. Part I" // https://arxiv.org/abs/2008.10509 (2020) [1].

The proposed in [1] physics-integrated (or PDE-integrated (partial differential equation)) machine learning (ML) framework is furtherly investigated. The Navier-Stokes equations are solved using the Tensorflow ML library for Python programming language via the Chorin's projection method. The Tensorflow solution is integrated with a deep feedforward neural network (DFNN). Such integration allows one to train a DFNN embedded in the Navier-Stokes equations without having the target (labeled training) data for the direct outputs from the DFNN; instead, the DFNN is trained on the field variables (quantities of interest), which are solutions for the Navier-Stokes equations (velocity and pressure fields).

To demonstrate performance of the framework, two additional case studies are formulated: 2D turbulent lid-driven cavities with predicted by a DFNN (a) turbulent viscosity and (b) derivatives of the Reynolds stresses. Despite its complexity and computational cost, the proposed physics-integrated ML shows a potential to develop a "PDE-integrated" closure relations for turbulent models and offers principal advantages, namely: (i) the target outputs (labeled training data) for a DFNN might be unknown and can be recovered using the knowledge base (PDEs); (ii) it is not necessary to extract and preprocess information (training targets) from big data, instead it can be extracted by PDEs; (iii) there is no need to employ a physics- or scale-separation assumptions to build a closure model for PDEs. The advantage (i) is demonstrated in the Part I paper [1], while the advantages (ii) and (iii) are the subjects for this study.

*Keywords*: Physics-integrated machine learning, PDE-integrated machine learning, physics-informed neural network, Navier-Stokes equations, ML embedded in PDE, closure model

## NOMENCLATURE

### Mathematical symbols

| | | | | |
|---|---|---|---|---|
| $\Delta t$ | Time step size, s | | $N_{data}$ | Number of datasets |
| $\varepsilon$ | Tolerance for convergence | | $p$ | Pressure, Pa |
| $\mu$ | Dynamic viscosity, Pa·s | | $t$ | Time, s |
| $\rho$ | Density, kg/m$^3$ | | $u$ | $x$-velocity, m/s |
| $A$ | Advection term, m/s$^2$ | | $v$ | $y$-velocity, m/s |
| $C$ | Cost (loss) function | | $x$ | Horizontal coordinate, m |
| $D$ | Diffusion term, kg/(m$^2$·s$^2$) | | $y$ | Vertical coordinate, m |
| $h$ | Mesh size, m | | $<\ >$ | Reynolds-averaged value |
| $k$ | Turbulence kinetic energy, m$^2$/s$^2$ | | | |

### Subscripts

| | | | | |
|---|---|---|---|---|
| $qss$ | Quasi steady state | | $sol$ | Solution |
| $i$ | Mesh node's $x$-index | | $targ$ | Target value |
| $inp$ | Input value | | $w$ | Wall |
| $j$ | Mesh node's $y$-index | | $x$ | $x$-projection of a vector |
| $nn$ | NN-based value | | $y$ | $y$-projection of a vector |

### Superscripts

| | | | | |
|---|---|---|---|---|
| $*$ | Predictor step (preliminary) value | | $n$ | Time step number |

### Dimensionless Numbers

| | |
|---|---|
| Re | Reynolds number |

### Acronyms

| | | | | |
|---|---|---|---|---|
| CNN | Convolutional Neural Network | | ML | Machine Learning |
| DD | Data-Driven | | NN | Neural Network |
| DFNN | Deep Feedforward Neural Network | | PDE | Partial Differential Equation |
| DNS | Direct Numerical Simulations | | RANS | Reynolds-Averaged Navier-Stokes |

1. INTRODUCTION

Incorporating and enforcing known flow physics is a challenge and opportunity for machine learning (ML) algorithms since most of the developed ML models are purely data-driven (DD) and, therefore, ignore the knowledge base accumulated throughout the centuries [2]. At the same time, traditional first principles fluid dynamics modelling (e.g. using the discretized Navier-Stokes equations) faces serious challenges associated with current equations-based analysis of fluids, including high dimensionality and nonlinearity, which defy closed-form solutions and limit real-time optimization and control efforts [4]. It is clear that the integration of the DD modelling with the knowledge base may tremendously improve and advance further development of the computational fluid dynamics.

This paper is investigation of an opportunity to integrate neural networks (NNs) with the numerical solution of partial differential equations (PDEs) (Type 3 ML framework [3]) and must be viewed as a continuation of the work performed in Part I paper [1]. Part I analyzes current ML and physics-informed ML approaches for fluid dynamics as well as demonstrates the opportunity and methodology for integration of NNs with the Navier-Stokes equations to build a surrogate model for non-constant velocity-dependent dynamic viscosity (case study No. 1). The methodology may be viewed as "bridge" between DD modelling and classical numerical methods for PDE solution. This paper furtherly investigates the proposed methodology and consider two additional case studies: 2D turbulent lid-driven cavities with NN-based (a) turbulent viscosity (case study No. 2) and (b) derivatives of the Reynolds stresses (case study No. 3). The obtained results confirm the advantages proposed in [1], namely: (i) the target outputs (labeled training data) for a deep feedforward neural network (DFNN) might be unknown and can be recovered using the knowledge base (PDEs); (ii) it is not necessary to extract and preprocess information (training targets) from big data, instead it can be extracted by PDEs; (iii) there is no need to employ a physics- or scale-separation assumptions to build a closure model for PDEs. The advantage (i) is demonstrated in Part I [1], the advantage (ii) is demonstrated in this paper (Part II), while item (iii) is the subject for future work.

Section 2 discusses the case study No. 2: building a surrogate model for turbulent viscosity and turbulence kinetic energy in 2D lid-driven cavity, including mathematical and computational models, data preprocessing and generation (including an attempt to generate data using the ANSYS Fluent fluid dynamics modelling software), physics-integrated ML architecture, and results. Similarly, Section 3 discusses the case study No. 3: building a surrogate model for derivatives of the Reynolds stresses for 2D lid-driven cavity problem. The obtained results show a potential to use PDE-integrated ML for developing surrogate closure relations for turbulence modelling.

## 2. CASE STUDY NO. 2: A SURROGATE MODEL FOR TURBULENT VISCOSITY

### 2.1. Mathematical and Computational Models for Data Generation

Considered mathematical model slightly differs from the model for the case study No. 1 in Part I [1] (a surrogate model for velocity-dependent dynamic viscosity). The main difference is that for the case study No. 2 the dynamic viscosity is constant, while the physical problem is the same – flow in 2D lid-driven cavity. Additionally, case study No. 2 will consider turbulent flow in the cavity, whereas case study No. 1 investigated laminar flow (low Re numbers).

Therefore, the governing equations are 2D incompressible Navier-Stokes equations with constant dynamic viscosity:

$$\nabla \cdot \vec{u} = 0 \tag{2.1.1}$$

$$\frac{\partial \vec{u}}{\partial t} + \nabla\left(\vec{u}\vec{u}^T\right) = -\frac{1}{\rho}\nabla p + \frac{\mu}{\rho}\nabla^2 \vec{u} \tag{2.1.2}$$

where $\vec{u}$ is velocity vector; $x$ and $y$ are horizontal and vertical coordinates, respectively; $t$ is time, $\rho$ is density, $\mu$ is dynamic viscosity, $p$ is pressure.

The boundary conditions are the same as for the case study No. 1 (no-slip impermeable walls; the upper one is moving with constant $x$-velocity $u_w$):

$$u(t, x = 0, y) = u(t, x = 1, y) = u(t, x, y = 0) = 0$$
$$u(t, x, y = 1) = u_w \tag{2.1.3}$$
$$v(t, x = 0, y) = v(t, x = 1, y) = v(t, x, y = 0) = v(t, x, y = 1) = 0$$

For the case study No. 2, the upper wall velocity is fixed: $u_w = 1$ m/s, while the dynamic viscosity is varied to obtain different Re numbers and generate training and validation data (on the contrary comparing to the case study No. 1, when $u_w$ varied to keep the dependency for dynamic viscosity constant).

The equations (2.1.1) and (2.1.2) are discretized on a uniform staggered grid with 102×102 control volumes with size $h$ and solved using the Chorin's projection method (similarly to the case study No. 1, but here two-step Adams-Bashforth method for time discretization is employed and the dynamic viscosity is constant (which simplify the numerical schemes – compare with Eqs. (2.1.3) – (2.1.5) in [1]):

$$u^*_{i+1/2,j} = u^n_{i+1/2,j} + \frac{3}{2}\Delta t\left[-(A_x)^n_{i+1/2,j} + \frac{1}{\rho}(D_x)^n_{i+1/2,j}\right] - \frac{1}{2}\Delta t\left[-(A_x)^{n-1}_{i+1/2,j} + \frac{\mu}{\rho}(D_x)^{n-1}_{i+1/2,j}\right]$$

$$v^*_{i,j+1/2} = v^n_{i,j+1/2} + \frac{3}{2}\Delta t\left[-(A_y)^n_{i,j+1/2} + \frac{1}{\rho}(D_y)^n_{i,j+1/2}\right] - \frac{1}{2}\Delta t\left[-(A_y)^{n-1}_{i,j+1/2} + \frac{\mu}{\rho}(D_y)^{n-1}_{i,j+1/2}\right]$$

(2.1.4)

$$\frac{p^{n+1}_{i+1,j} + p^{n+1}_{i-1,j} + p^{n+1}_{i,j+1} + p^{n+1}_{i,j-1} - 4p^{n+1}_{i,j}}{h^2} = \frac{\rho}{\Delta t}\left(\frac{u^*_{i+1/2,j} - u^*_{i-1/2,j} + v^*_{i,j+1/2} - v^*_{i,j-1/2}}{h}\right) \quad (2.1.5)$$

$$u^{n+1}_{i+1/2,j} = u^*_{i+1/2,j} - \frac{\Delta t}{\rho h}\left(p^{n+1}_{i+1,j} - p^{n+1}_{i,j}\right)$$

$$v^{n+1}_{i,j+1/2} = v^*_{i,j+1/2} - \frac{\Delta t}{\rho h}\left(p^{n+1}_{i,j+1} - p^{n+1}_{i,j}\right)$$

(2.1.6)

where $n$ denotes number of a time step, $\Delta t$ is time step size, $A$ and $D$ are advection and diffusion terms, respectively; subscripts $i$ and $j$ denote numbers of control volumes, superscript * denotes preliminary value for velocities (predictor step).

The discretizations of the advection and diffusion terms are obtained using the finite volume method (compare with Eq. (2.1.6) and (2.1.7) in [1]):

$$(A_x)^n_{i+1/2,j} = \frac{1}{4h}\left\{\left(u^n_{i+3/2,j} + u^n_{i+1/2,j}\right)^2 - \left(u^n_{i+1/2,j} + u^n_{i-1/2,j}\right)^2 + \right.$$

$$\left. + \left(u^n_{i+1/2,j+1} + u^n_{i+1/2,j}\right)\left(v^n_{i+1,j+1/2} + v^n_{i,j+1/2}\right) - \left(u^n_{i+1/2,j} + u^n_{i+1/2,j-1}\right)\left(v^n_{i+1,j-1/2} + v^n_{i,j-1/2}\right)\right\}$$

$$(A_y)^n_{i,j+1/2} = \frac{1}{4h}\left\{\left(v^n_{i,j+3/2} + v^n_{i,j+1/2}\right)^2 - \left(v^n_{i,j+1/2} + v^n_{i,j-1/2}\right)^2 + \right.$$

$$\left. + \left(v^n_{i+1,j+1/2} + v^n_{i,j+1/2}\right)\left(u^n_{i+1/2,j+1} + u^n_{i+1/2,j}\right) - \left(v^n_{i,j+1/2} + v^n_{i-1,j+1/2}\right)\left(u^n_{i-1/2,j+1} + u^n_{i-1/2,j}\right)\right\}$$

(2.1.7)

$$(D_x)^n_{i+1/2,j} = \frac{1}{h^2}\left(u^n_{i+3/2,j} + u^n_{i+1/2,j+1} + u^n_{i-1/2,j} + u^n_{i+1/2,j-1} - 4u^n_{i+1/2,j}\right)$$

$$(D_y)^n_{i,j+1/2} = \frac{1}{h^2}\left(v^n_{i,j+3/2} + v^n_{i+1,j+1/2} + v^n_{i,j-1/2} + v^n_{i-1,j+1/2} - 4v^n_{i,j+1/2}\right)$$

(2.1.8)

The described above computational model with boundary conditions Eq. (2.1.3) is programmed in Fortran and the code is used to generate training and validation data for the ML model. The next subsection discusses the data generation and preprocessing activities

## 2.2. Training and Validation Data Parameters

Since the turbulent flow is considered, it is necessary to perform time-averaging for pressure and velocity fields before using them in ML models. For that a following methodology was employed. When an infinity norm

$$L_{vel} = \max (|u^n - u^{n-1}|, |v^n - v^{n-1}|) \qquad (2.2.1)$$

reaches a value $\varepsilon_{qss} = 5 \cdot 10^{-5}$ m/s for quasi-steady state and the number of time iterations is greater than $10^4$ (these thresholds were picked manually), a simulation is continued for $50 \times 10^3$ time steps, while the fields are being written to the output file every 500 time steps. Therefore, each training/validation dataset produces 101 pressure and velocity fields. Then these values are time averaged (since the time step and mesh discretization are constant, just average value is to be found in each control volume). In the end, each training/validation dataset will provide one averaged pressure, one averaged $x$- and one averaged $y$-velocity fields. Overall 21 training and 3 validation datasets are generated according to Table 2.2.1.

Fig. 2.2.1 shows the dependence of a norm $L_{vel}$ (Eq. (2.2.1)) during time iterations for the training dataset No. 1 on a quasi-steady state regime (after $10^4$ time steps); i.e. the averaging was performed in a shown time frame.

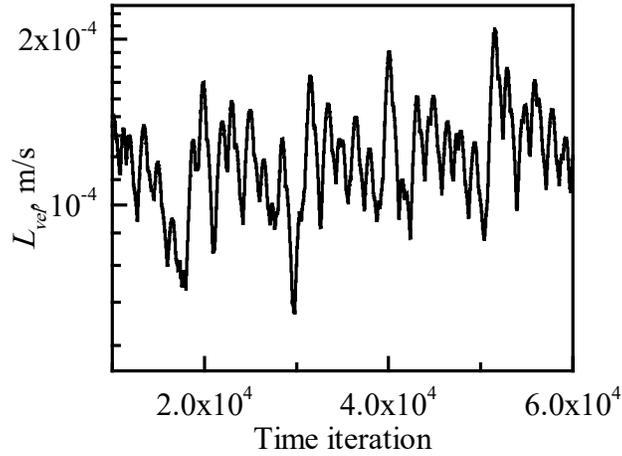

Fig. 2.2.1. Velocity norm dependence during the iterations for the training dataset No. 1 (during this range of time steps the fields are being saved for the averaging).

Table 2.2.1. Generated data parameters.

| No. | Re | μ×10³, Pa·s | No. | Re | μ×10³, Pa·s | No. | Re | μ×10³, Pa·s |
|---|---|---|---|---|---|---|---|---|
| **Training Datasets** | | | | | | **Validation Datasets** | | |
| 1 | 10000 | 0.10000 | 12 | 12200 | 0.08197 | 1 | 9800 | 0.10204 |
| 2 | 10200 | 0.09804 | 13 | 12399 | 0.08065 | 2 | 12100 | 0.08264 |
| 3 | 10400 | 0.09615 | 14 | 12599 | 0.07937 | 3 | 14200 | 0.07042 |
| 4 | 10600 | 0.09434 | 15 | 12799 | 0.07813 | | | |
| 5 | 10800 | 0.09259 | 16 | 13001 | 0.07692 | | | |
| 6 | 11000 | 0.09091 | 17 | 13200 | 0.07576 | | | |
| 7 | 11199 | 0.08929 | 18 | 13399 | 0.07463 | | | |
| 8 | 11400 | 0.08772 | 19 | 13600 | 0.07353 | | | |
| 9 | 11600 | 0.08621 | 20 | 13801 | 0.07246 | | | |
| 10 | 11799 | 0.08475 | 21 | 14000 | 0.07143 | | | |
| 11 | 12000 | 0.08333 | | | | | | |

Fig. 2.2.2 compares the averaged fields with randomly picked instantaneous fields (output No. 27/101) for the training dataset No. 1. There are some minor differences between the averaged and instantaneous fields due to the turbulent nature of the flow, which are almost no noticeable. Since the case study No. 2 is aimed at developing a closure model for turbulent viscosity, the Re number is varied near the value $12 \cdot 10^3$ (see Table 2.2.1). Therefore, the (time-averaged) steady state solutions are very different from the "laminar" solutions (case study No. 1, see Fig. 4.1.1 in [1]).

Fig. 2.2.3 demonstrates averaged streamtraces in the domain. It is clearly seen that the flow is much more complex comparing to the laminar flow: there are 4-5 vortices and they are larger and have more complex structures (compare with Fig. 4.1.1 in [1]).

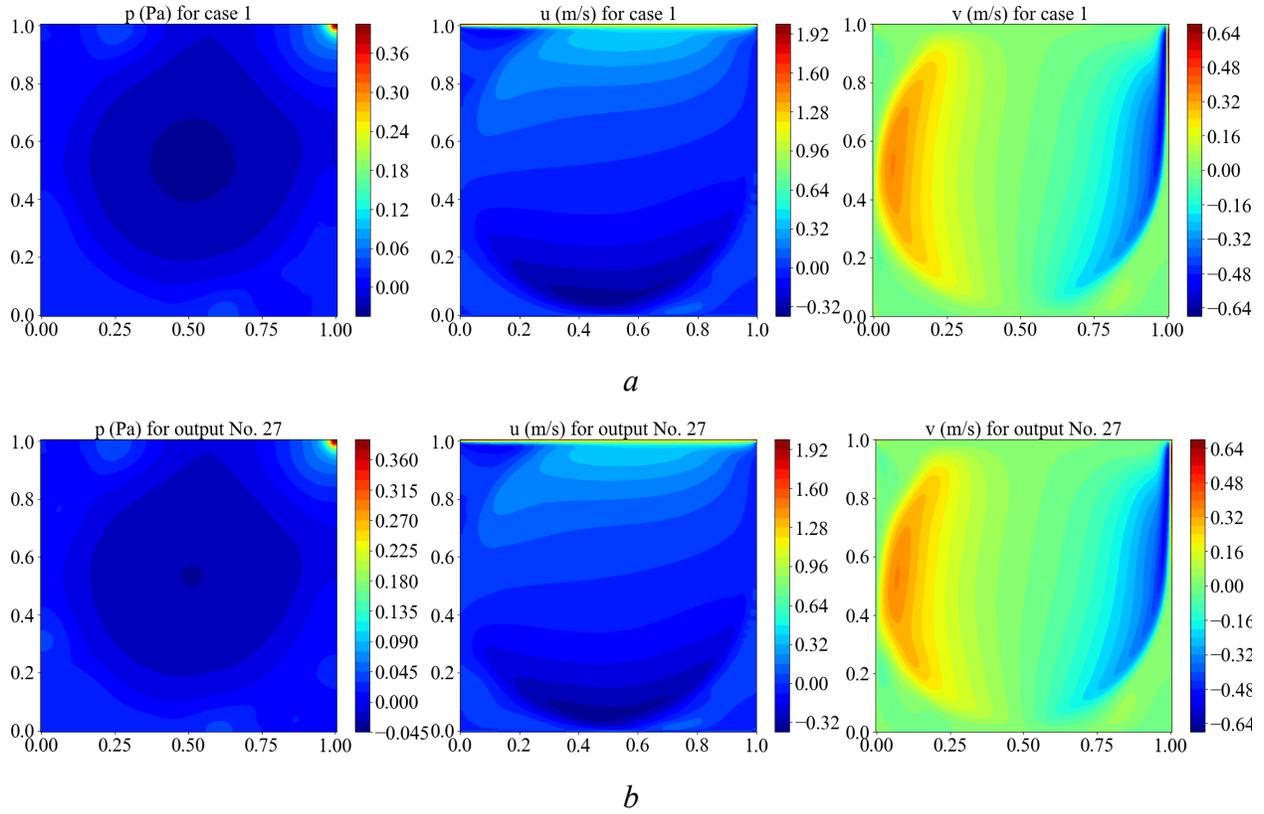

Fig. 2.2.2. Steady state solutions for 2D turbulent lid-driven cavity for a training dataset No. 1 (Re = $10^4$): (*a*) averaged pressure, *x*-velocity, *y*-velocity; (*b*) instantaneous fields (output No. 27/101).

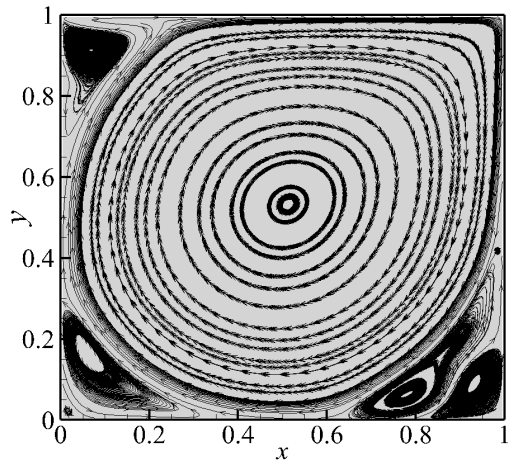

Fig. 2.2.3. Time-averaged streamtraces in the domain for a training dataset No. 1 (Re = $10^4$).

## 2.3. On the Attempt to Generate Data Using ANSYS Fluent

There was also an attempt to generate training/validation data using the ANSYS Fluent fluid simulation software. Different turbulent models were considered including *k-ε*, *k-ω*, and *k-ω* SST models (the last two gave very similar results). The profiles along centerlines of the cavity for Re = $10^4$ (training dataset No. 1) are presented on Fig. 2.3.1 together with Direct Numerical Simulations (DNS[1]) time-averaged data (Fortran calculations) and reference values from by Ghia U. et al. (1982) [6].

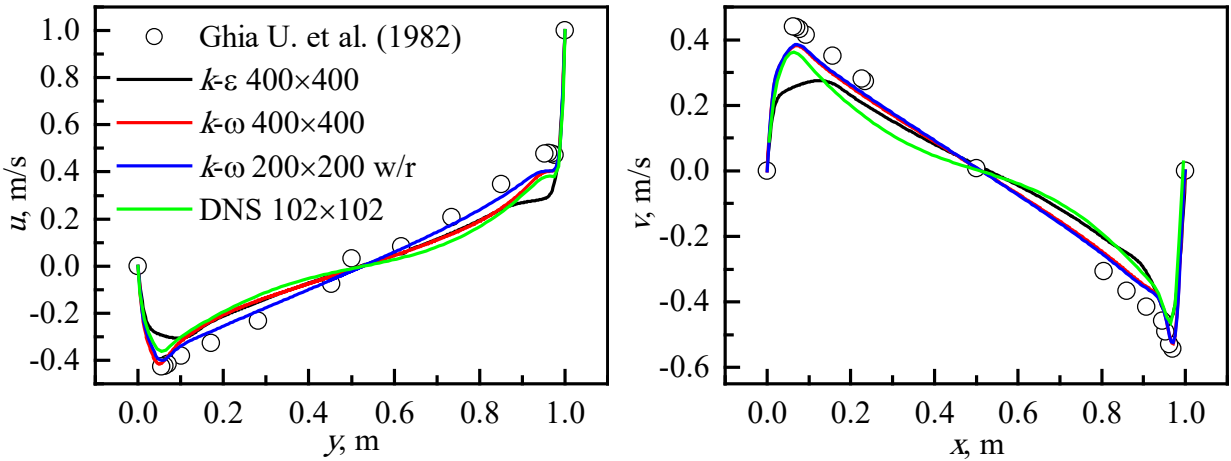

Fig. 2.3.1. Velocity profiles along centerlines of the cavity for Re = $10^4$.

From Fig. 2.3.1 one can make following conclusions about the data generated using the ANSYS Fluent:

- *k-ε* model shows worse performance for this problem comparing to *k-ω* model (as expected, since *k-ε* model works well for external flows, while *k-ω* model works well for internal flows) [5];
- mesh convergence study for *k-ε* model shows that 400×400 uniform mesh is enough for grid convergence (control volume size $h \sim 2.5 \cdot 10^{-3}$ m);
- mesh convergence study for *k-ω* model shows that 400×400 uniform mesh is enough for grid convergence (control volume size $h \sim 2.5 \cdot 10^{-3}$ m);
- both turbulent models underpredict velocity profiles comparing to reference values [6].

---

[1] Usually, DNS implies finer meshes and 3D modelling for turbulence. Here we use "DNS" to denote direct solution for the Navier-Stokes equations (without using closures for turbulence).

Such meshes (400×400) are too fine for solving the Navier-Stokes equations using the Tensorflow library [7]. Therefore, a non-uniform mesh (200×200) with refinement (w/r) near the walls was generated and simulation performed using the *k-ω* model. The obtained results are close to the simulation results for the mesh 400×400 (see Fig. 2.3.1).

For such non-uniform mesh, the training and validation data were generated according to Table 2.2.1. Then this data was mapped into a uniform mesh (32×32, 102×102, 202×202) using the nearest neighbor algorithm available through SciPy library [8] to reduce the computational cost for the ML model (the developed ML model can only work with a uniform grid). When this mapped data was used for training, a neural network (NN) was not be able to catch the underlying dependency. Therefore, the mapped data cannot be used for training of Type 3 ML probably since the spatial structure is being lost during the mapping procedure. Additionally, the ANSYS Fluent does not allow to generate ideally uniform mesh and the data mapping will be always needed. Therefore, the decision was made to generate data using DNS according to the Section 2.2. 102×102 mesh is used (Fig. 2.3.1). Surely, the mesh refinement is not enough, but the computational cost for such mesh is acceptable and numerical error ("numerical viscosity") is sufficiently small for demonstration purposes.

## 2.4. PDE-integrated ML Architecture

The developed architecture for Type 3 ML for the case study No. 2 is presented on Fig. 2.4.1.

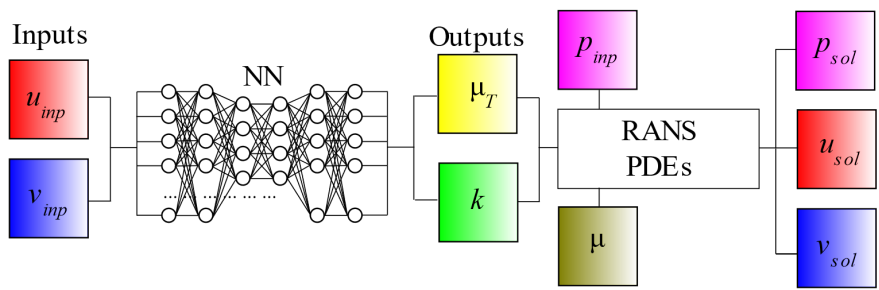

Fig. 2.4.1. Type 3 architecture for the case study No. 2.

As inputs to the NN steady state solutions of the Navier-Stokes equations (averaged DNS data) are used: $u_{inp}$ and $v_{inp}$. Steady state pressure data $p_{inp}$ is also required to be used in the Reynolds-Averaged Navier-Stokes (RANS) block as well as constant dynamic viscosities μ (see

Table 2.2.1). The NN is trained to predict turbulent (eddy) viscosity $\mu_T$ and turbulence kinetic energy $k$ fields, which are then used to solve the RANS equations with employed turbulent viscosity hypothesis and predict velocity and pressure fields $u_{sol}$, $v_{sol}$, and $p_{sol}$. These solutions are then used in the cost function

$$C = \frac{1}{2} \frac{\sum\left(u_{sol_{i,j}} - u_{targ_{i,j}}\right)^2 + \sum\left(v_{sol_{i,j}} - v_{targ_{i,j}}\right)^2 + \sum\left(p_{sol_{i,j}} - p_{targ_{i,j}}\right)^2}{N_{data}} \qquad (2.4.1)$$

where target values are the same with inputs $u_{targ} = u_{inp}$, $v_{targ} = v_{inp}$, $p_{targ} = p_{inp}$ (autoencoder-like architecture), $N_{data}$ is the number of datasets.

The main distinguishing feature comparing to the case study No. 1 is that pressure fields are also used in the cost function. This greatly improved the results and made the NN predictions more stable for convergence of the solutions to steady state (however, for the case study No. 1 the performance was not improved by adding pressure in the cost function). We also note that the convergence to steady state is only possible from close to steady state initial conditions. This is because the NN is trained only on steady-state data. Probably, more sophisticated architectures are needed to capture time-dependent mappings (e.g. Recurrent Neural Networks [10] with originally proposed in [3] Type 3 framework with global iterations [1]).

The RANS equations (2.4.2) and (2.4.3) are the results of the Reynolds-averaging of the Navier-Stokes equations:

$$\frac{\partial u_i}{\partial x_i} = 0$$

$$\frac{\partial u_i}{\partial t} + u_j \frac{\partial u_i}{\partial x_j} = -\frac{1}{\rho}\frac{\partial p}{\partial x_i} + \frac{1}{\rho}\frac{\partial}{\partial x_j}\left(\mu\left(\frac{\partial u_i}{\partial x_j} + \frac{\partial u_j}{\partial x_i}\right)\right)$$

$$u_i = U_i + u_i'$$

$$p = P + p'$$

$$\frac{\partial U_i}{\partial x_i} = 0 \qquad (2.4.2)$$

$$\frac{\partial U_i}{\partial t} + U_j \frac{\partial U_i}{\partial x_j} = -\frac{1}{\rho}\frac{\partial P}{\partial x_i} + \frac{1}{\rho}\frac{\partial}{\partial x_j}\left(\mu\left(\frac{\partial U_i}{\partial x_j} + \frac{\partial U_j}{\partial x_i}\right) - \rho\langle u'_i u'_j \rangle\right) \qquad (2.4.3)$$

where capital letters denote averaged components, primed letter denote fluctuation components.

According to the turbulent viscosity hypothesis the Reynolds stresses $\langle u'_i u'_j \rangle$ are proportional to the mean rate of strain [9]:

$$-\rho \langle u'_i u'_j \rangle + \frac{2}{3}\rho k \delta_{ij} = \mu_T \left( \frac{\partial U_i}{\partial x_j} + \frac{\partial U_j}{\partial x_i} \right) \tag{2.4.4}$$

and Eq. (2.2.3) becomes

$$\frac{\partial U_i}{\partial t} + U_j \frac{\partial U_i}{\partial x_j} = -\frac{1}{\rho}\frac{\partial P}{\partial x_i} + \frac{1}{\rho}\frac{\partial}{\partial x_j}\left[(\mu+\mu_T)\left(\frac{\partial U_i}{\partial x_j}+\frac{\partial U_j}{\partial x_i}\right)\right] - \frac{2}{3}\frac{\partial k}{\partial x_i} \tag{2.4.5}$$

where often dynamic and turbulent viscosities are written as effective viscosity $\mu_{eff} = \mu + \mu_T$.

The equations (2.4.2) and (2.4.5) are solved using the Tensorflow ML library via the Chorin's projection method by employing similar methodology as for the case study No. 1 (see Eqs. (2.2.3) – (2.2.6) in [1]), except modification to diffusion terms:

$$(D_x)^n = \frac{1}{h^2}\left\{ 2 \cdot \ker_2\left[\mu_{eff}^n \cdot \ker_2(u^n)\right] + \ker_3\left[\ker_4(\mu_{eff}^n) \cdot (\ker_3(u^n) + \ker_2(v^n))\right] - \frac{2\rho h}{3}\ker_2(k) \right\}$$

$$(D_y)^n = \frac{1}{h^2}\left\{ 2 \cdot \ker_3\left[\mu_{eff}^n \cdot \ker_3(v^n)\right] + \ker_2\left[\ker_4(\mu_{eff}^n) \cdot (\ker_2(v^n) + \ker_3(u^n))\right] - \frac{2\rho h}{3}\ker_3(k) \right\} \tag{2.4.6}$$

A DFNN is used in this study, which has the input layer with $101 \times 102 + 102 \times 101 = 20604$ neurons (fed with $u_{inp}$ and $v_{inp}$ fields), 4 hidden layers with 30 neurons in each layer, and the output layer with $100 \times 100 + 100 \times 100 = 20000$ neurons, which predicts turbulent dynamic viscosity and turbulence kinetic energy fields in the centers of control volumes. The other hyperparameters are the same as in [1]. The DFNN was trained for 15,000 epochs. A convolutional NN (CNN) was also used and gave a similar with the DFNN performance.

### 2.5. Results

After the training, the DFNN was used to predict turbulent viscosity and turbulence kinetic energy for the training and validation data. *Out of curiosity*, they are compared to the values obtained from the ANSYS Fluent calculations ($k$-$\omega$ SST model, mesh $200 \times 200$ with refinement), Fig. 2.5.1. As it could be seen, the turbulent viscosities have similar qualitative distribution (large values in center of the cavity and lower values in the periphery). However, ANSYS predicts maximum values in the center of the domain, while turbulent viscosity extracted from the DNS data has maximum values shifted to the right bottom corner. Turbulence kinetic energies have similar qualitative and quantitative behavior.

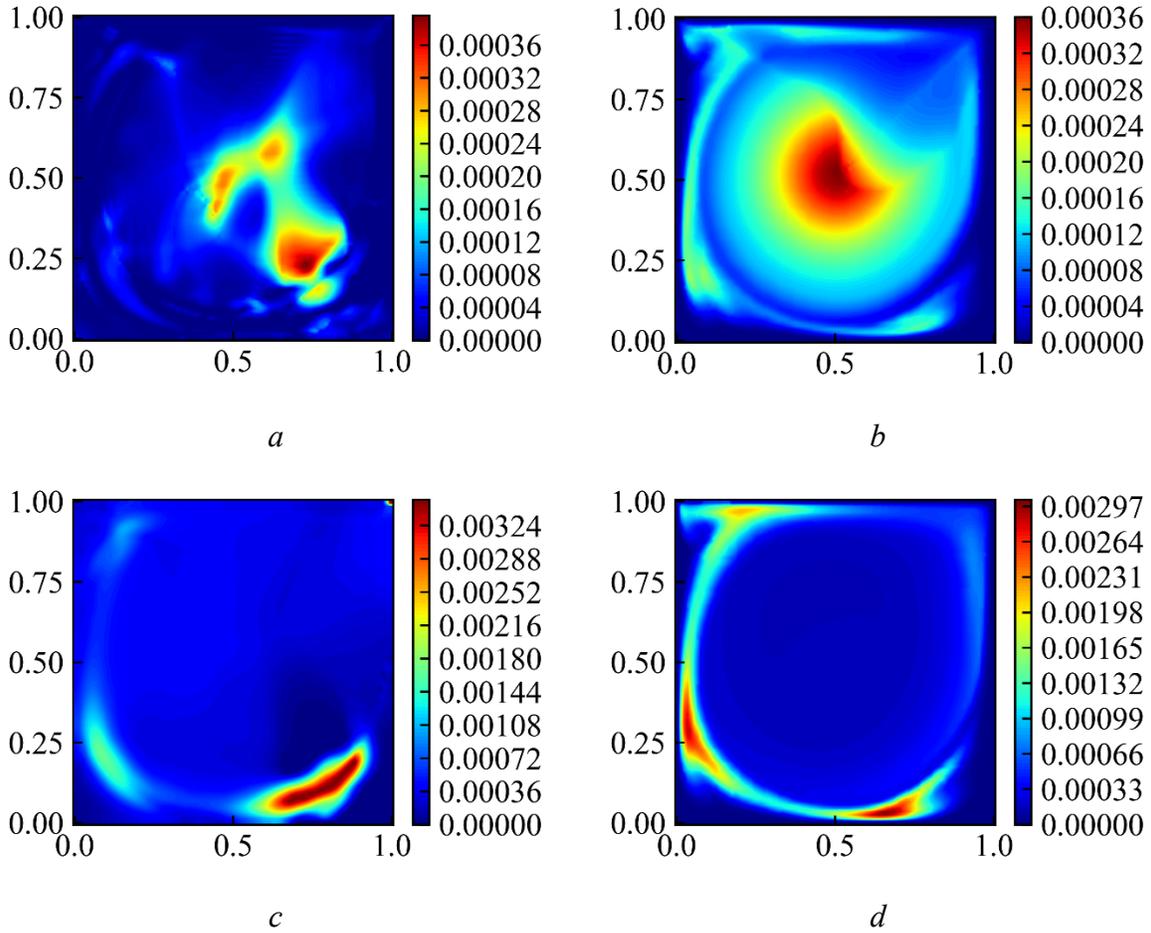

Fig. 2.5.1. Turbulent viscosity (Pa·s) fields for validation dataset No. 1: (*a*) extracted by the DFNN, (*b*) ANSYS Fluent data. Turbulence kinetic energy (m$^2$/s$^2$) fields for validation dataset No. 1: (*c*) extracted by the DFNN, (*d*) ANSYS Fluent data.

Fig. 2.5.2 demonstrates the deviations of the solutions obtained using the NN-based turbulent viscosity for the validation dataset No. 3 (worst obtained results). As could be seen, the recovered turbulent viscosity fields from the NN allow to obtain solutions that are very close to the exact ones.

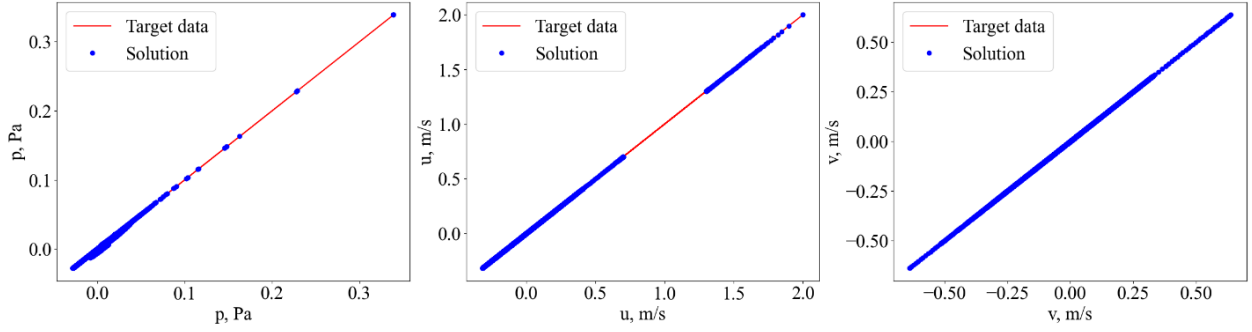

Fig. 2.5.2. Deviations of the solutions with the NN-based turbulent viscosity from the exact values for the validation dataset No. 3.

## 3. CASE STUDY NO. 3: A SURROGATE MODEL FOR DERIVATIVES OF REYNOLDS STRESSES

### 3.1. PDE-integrated ML Architecture

The same data are used to explore the case study No. 3. However, the turbulent hypothesis is not adopted (Eq. (2.4.3) is solved using the Tensorflow instead of Eq. (2.4.5)):

$$\frac{\partial U_i}{\partial t} + U_j \frac{\partial U_i}{\partial x_j} = -\frac{1}{\rho}\frac{\partial P}{\partial x_i} + \frac{\mu}{\rho}\frac{\partial^2 U_i}{\partial x_j \partial x_j} - \frac{\partial \langle u'_i u'_j \rangle}{\partial x_j} \quad (3.1.1)$$

Since in the Eq. (3.1.1) dynamic viscosity is constant and there are additional terms (Reynolds stresses derivatives) in the momentum equations, the solutions methodology on the Tensorflow (diffusion terms and predictor step) should be modified as follows (compare to Eqs. (2.2.4) in [1]):

$$(D_x)^n = \frac{\mu}{h^2}\left\{2\cdot\ker_2\left[\ker_2(u^n)\right] + \ker_3\left[\ker_3(u^n) + \ker_2(v^n)\right]\right\}$$

$$(D_y)^n = \frac{\mu}{h^2}\left\{2\cdot\ker_3\left[\ker_3(v^n)\right] + \ker_2\left[\left(\ker_2(v^n) + \ker_3(u^n)\right)\right]\right\}$$

$$u^* = u^n + \Delta t\left[-(A_x)^n + \frac{1}{\rho}(D_x)^n - \left(\frac{\partial\langle u'u'\rangle}{\partial x} + \frac{\partial\langle u'v'\rangle}{\partial y}\right)_{nn}\right]$$

$$v^* = v^n + \Delta t\left[-(A_y)^n + \frac{1}{\rho}(D_y)^n - \left(\frac{\partial\langle v'v'\rangle}{\partial y} + \frac{\partial\langle v'u'\rangle}{\partial x}\right)_{nn}\right]$$

where the subscript *nn* means that these values are directly predicted by a NN (without discretization).

The developed architecture for Type 3 ML for the case study No. 3 is presented on Fig. 3.1.1. Here the NN predicts the derivatives for the Reynolds stresses, which are then used to obtain the solutions for pressure and velocity fields. The solutions are then compared with the target data using the cost function Eq. (2.4.1).

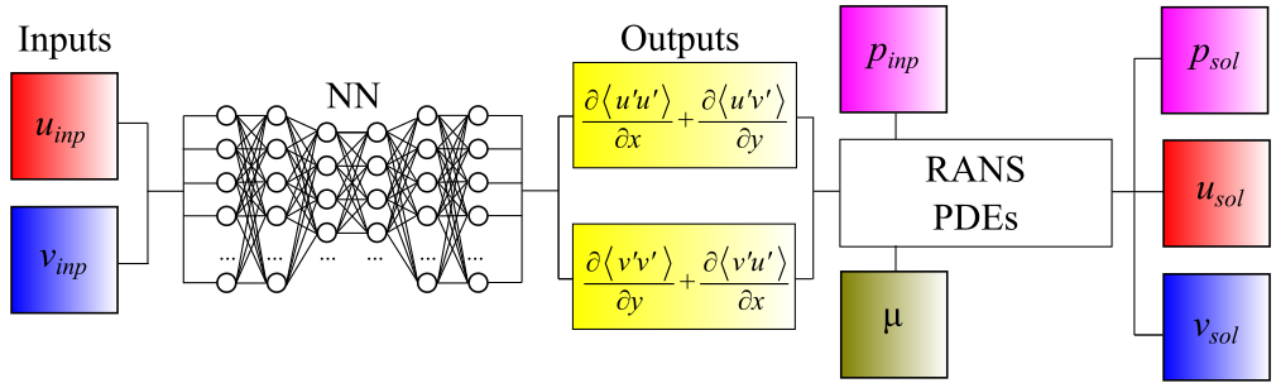

Fig. 3.1.1. Type 3 architecture for the case study No. 3.

A similar DFNN is used for this case study, which has the input layer with 101×102 + 102×101 = 20604 neurons (fed with $u_{inp}$ and $v_{inp}$ fields), 4 hidden layers with 128 neurons in each layer, and the output layer with 101×102 + 102×101 = 20604 neurons to predict the derivatives of the Reynolds stresses on edges of control volumes. The DFNN was trained for 30,000 epochs. The other hyperparameters are the same as in [1], except the employed dropout technique (25%) to prevent overfitting. A CNN was also used and gave similar with the DFNN performance.

### 3.2. Results

Fig. 3.2.1 demonstrates the fields of Reynolds stresses derivatives recovered by the DFNN for the validation dataset No. 1. The maximum values of the derivatives are located in the right bottom corner, which agrees with Fig. 2.5.1.

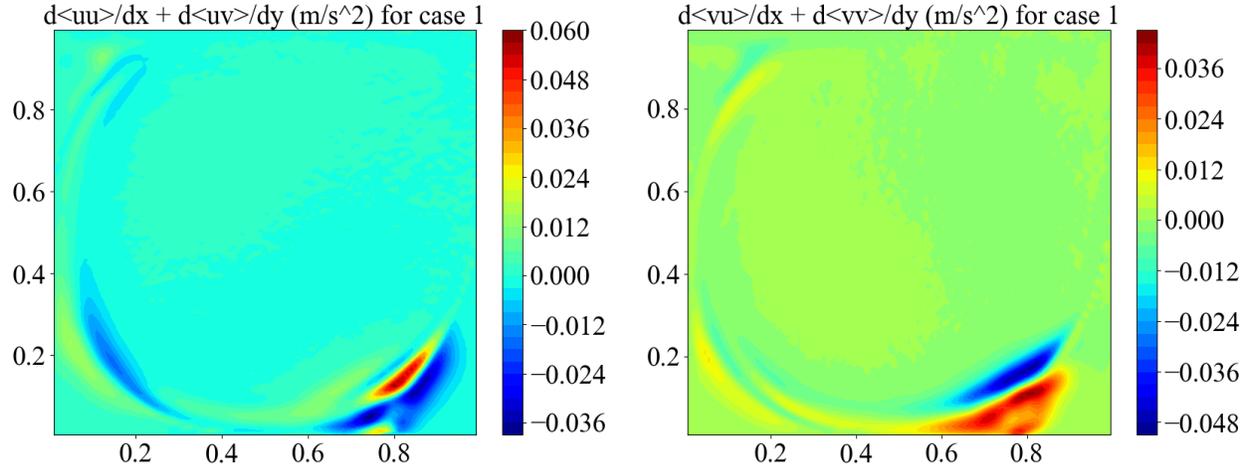

Fig. 3.2.1. Reynolds stresses derivatives returned by the DFNN for the validation dataset No. 1.

Fig. 2.2.2 demonstrates the deviations of the solutions obtained using the NN-based turbulent viscosity for the validation dataset No. 3 (worst obtained results). As could be seen, the obtained steady-state solution is worse than for the case study No. 2 (Fig. 2.5.2).

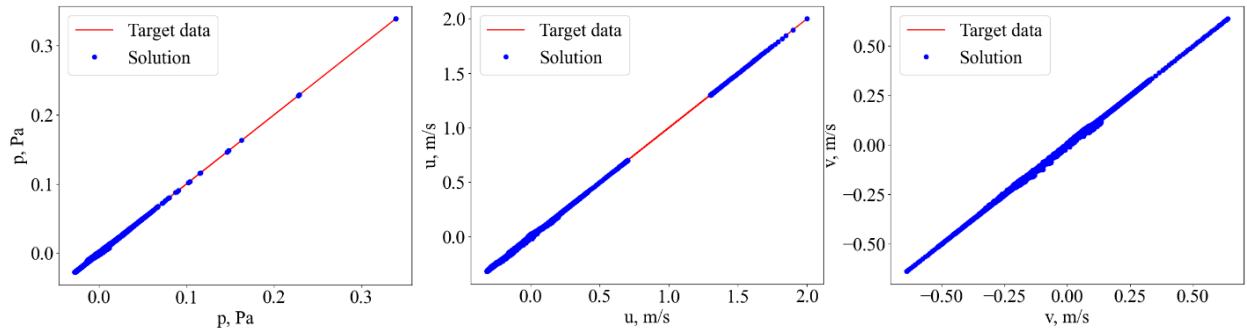

Fig. 2.2.2. Deviations of the solutions with the NN-based derivatives of Reynolds stresses from the exact values for the validation dataset No. 3.

## 4. CONCLUSIONS

This work is a continuation of the paper [1], where physics-integrated ML framework (Type 3) is proposed to predict non-constant velocity-dependent dynamic viscosity fields in the Navier-Stokes equations (case study No. 1). As discussed in [1], the developed framework is promising because (i) it allows to recover unknown physical values from the field variables if the governing equations for physics are known; (ii) it eliminates the necessity to extract physically-interpretable data from big data to train a NN; (iii) it eliminates the need to postulate a scale and physics separation. While the case study No. 1 explicitly demonstrates the item (i) – unknown velocity-dependent dynamic viscosity is recovered by PDEs, case studies No. 2 and 3 demonstrate items (ii) and (iii), respectively – turbulent viscosity, turbulence kinetic energy and derivatives of Reynolds stresses directly extracted from the velocity fields by a NN, without the necessity to extract them manually (as it usually done). Additionally, the physics-integrated framework is flexible to be switched between different ways of building closures (e.g. turbulent viscosity, or derivatives of the Reynolds stresses) without manipulations with "big" data. Type 3 is able to predict the net effect of different terms in PDEs ($\partial \langle u'u' \rangle / \partial x + \partial \langle u'v' \rangle \partial y$ or $\partial \langle v'v' \rangle / \partial y + \partial \langle u'v' \rangle \partial y$), without separating them. Even though is case study No. 3 these terms represented the same physics, in general (for more complex case studies), these terms may represent different physics (or scales). For example, it may be the net force on a bubble in a multiphase flow.

Thus, considered case studies demonstrate a potential for building closure relations for turbulence modelling with direct employment of DNS data. However, there are several challenges that need to be tackled in the future work, e.g. capturing the time-dependent mappings (recurrent neural networks may be useful for such problems); usage of data from different solvers, when discretization schemes and meshes are different in a data generation tool and in ML framework; slow performance of Python and ML libraries for numerical solution of PDEs (as a result, high computational cost for 3D problems); usage of the trained NNs for different meshes and problems; solution of the Navier-Stokes equation using the Tensorflow on unstructured grid.

**ACKNOWLEDGEMENTS**

This research was performed with support of North Carolina State University Provost Doctoral Fellowship to the first author. The authors are also grateful to Dr. Chih-Wei Chang for his guidance in the implementation of the Type 3 ML framework.**REFERENCES**

1. Iskhakov A.S., Dinh N.T. Physics-integrated machine learning: embedding a neural network in the Navier-Stokes equations // https://arxiv.org/abs/2008.10509, 2020
2. Willard J., Jia X, Xu S., Steinbach M., Kumar V. Integrating physics-based modeling with machine learning: a survey // https://arxiv.org/abs/2003.04919, 2020